\documentclass[twocolumn,showpacs,preprintnumbers,amsmath,amssymb,superscriptaddress]{revtex4}
\usepackage{graphicx}
\usepackage{amsmath}
\usepackage{amssymb}
\usepackage{dcolumn}
\usepackage{bm}
\usepackage{natbib}
\providecommand{\norm}[1]{\left\vert #1\right\vert}
\providecommand{\abs}[1]{\norm{#1}}%

\providecommand{\ket}[1]{\left\vert #1\right\rangle}
\providecommand{\braket}[2]{\left\langle #1 | #2 \right\rangle}

\providecommand{\Tp}{\ensuremath{\ket{\mathrm{T}\ensuremath{^+}}}} %
\providecommand{\s}{\ensuremath{S(0,2)}}%
\providecommand{\hf}{\ensuremath{A}}%
\providecommand{\hftot}{\ensuremath{\hf}}%

\providecommand{\makeket}[4]{\ket{\text{#1}_{#2} \text{#3}_{#4}}}%
\providecommand{\rrket}{\makeket{R}{\uparrow}{R}{\downarrow}}%
\providecommand{\bket}{\makeket{L}{\uparrow}{R}{\downarrow}}%
\providecommand{\cket}{\makeket{L}{\downarrow}{R}{\uparrow}}%
\providecommand{\tket}{\makeket{L}{\uparrow}{R}{\uparrow}}%
\providecommand{\Pket}{\ket{\Psi}}
\providecommand{\LR}{\braket{L}{R}}%
\newcommand{\ua}{\uparrow}
\newcommand{\da}{\downarrow}
\newcommand{\br}{{\bf r}}

\begin{document}

\title{Inhomogeneous Nuclear Spin Flips}

\author{M. Stopa}
\email{stopa@cns.fas.harvard.edu} \affiliation{Center for Nanoscale
Systems, Harvard University, Cambridge, MA 02138}

\author{J. J. Krich}
\affiliation{Department of Physics, Harvard University, Cambridge,
MA 02138}

\author{A. Yacoby}
\affiliation{Department of Physics, Harvard University, Cambridge,
MA 02138}

\begin{abstract}
We discuss a feedback mechanism between electronic states in a
double quantum dot and the underlying nuclear spin bath. We analyze
two pumping cycles for which this feedback provides a force for the
Overhauser fields of the two dots to either equilibrate or diverge.
Which of these effects is favored depends on the g-factor and
Overhauser coupling constant $\hftot$ of the material. The strength
of the effect increases with $\hftot/V_x$, where $V_x$ is the
exchange matrix element, and also increases as the external magnetic
field $B_{ext}$ decreases.
\end{abstract}

\maketitle


Hyperfine interaction with the host nuclei in nanoscale GaAs
systems, while relatively weak, can nevertheless limit the electron
coherence time and thereby complicate strategies to implement
quantum information and quantum computing schemes in these systems
\cite{Marcus,Ono04,DasSarma06,Loss98}. Conversely, ever-increasing
control of angular momentum transfer between electrons and nuclei in
a range of materials enables numerous applications precisely because
of the environmental isolation of the nuclear system. These include
applications to quantum information processing employing NMR
\cite{Vandersypen04}. From the perspective of fundamental physics,
experiments on few-electron systems with controllable coupling to
the nuclear many-body system uncover a fascinating arena of new
phenomena with ramifications for theoretical physics and engineering
\cite{Hanson07}.

Experiments on double quantum dots with electron number $N=2$ have
uncovered and exploited an intriguing phenomenon called the ``Pauli
blockade" \cite{Ono02} in which two electrons with parallel spins
are forbidden from combining in one dot by the exclusion principal.
In transport or in gate pulsing, even when such a transition becomes
energetically favorable, it can only proceed via a spin ``flip-flop"
process in which angular momentum is exchanged with the local
nuclei. Repeating the spin transfer modifies the character of the
nuclear spin distribution. One metric for the nuclear state is the
difference between the total Overhauser fields of each dot. These
are the effective Zeeman fields which the electrons experience due
to nuclear polarization. Several recent experiments addressed
transfer of angular momentum from the electron system to the nuclear
bath through various pumping cycles. One experiment claims that
under a specific, repeated pulsing sequence (see below)
\cite{Reilly08} the polarizations in the two dots tend to
equilibrate; a phenomenon which has been numerically reproduced
\cite{Ramon_Hu07,Ribeiro08}. However, another similar experiment
claims to find a large difference induced between the Overhauser
fields of the two dots \cite{Foletti09}. The theory which we
describe here does not claim to explain either experiment.

\begin{figure}
\begin{center}
\includegraphics[bb=0.0in 1.0in 11.0in 11.0in,width=3.375in]{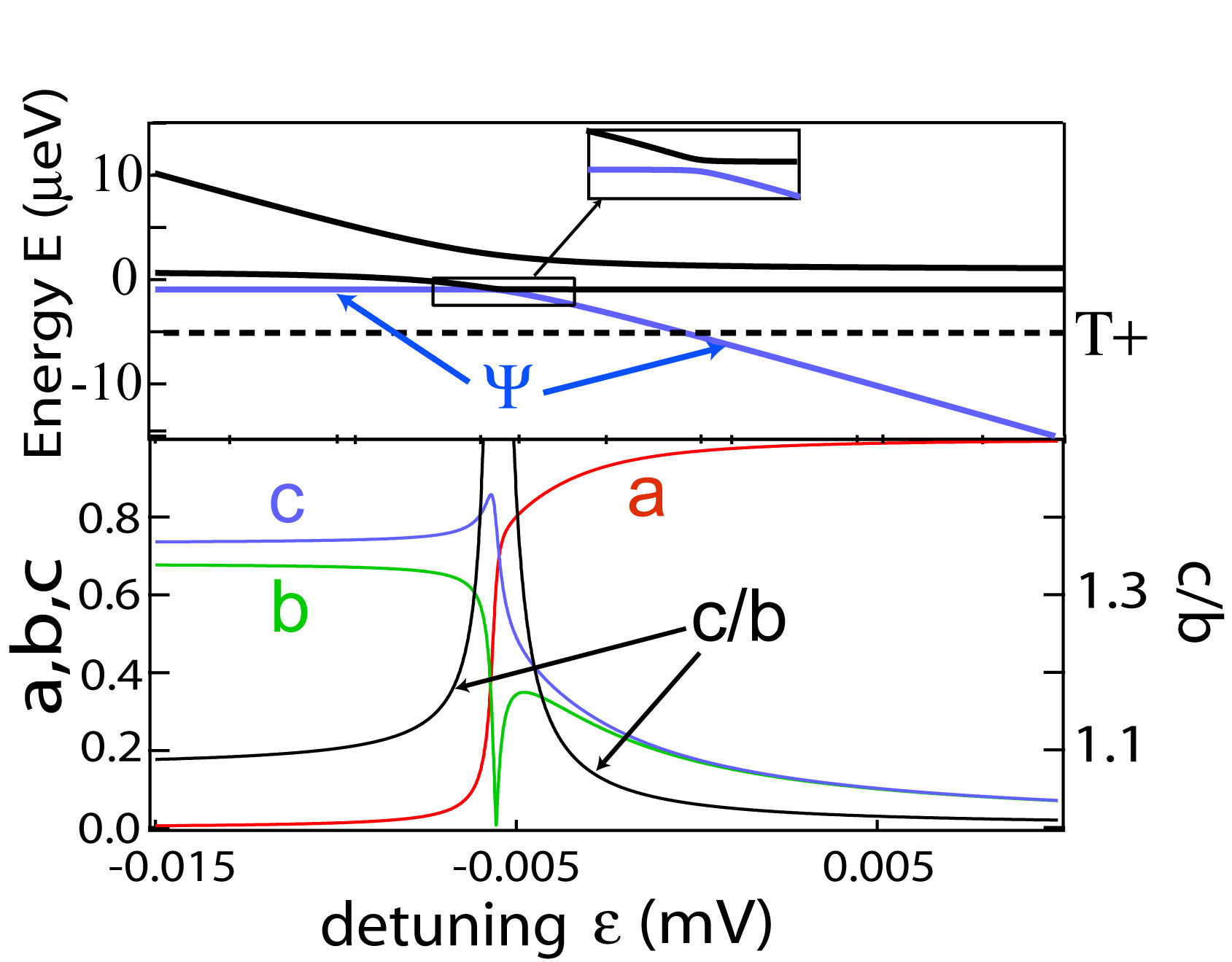}
\end{center}
\caption{Top: electronic states near the (1,1) to (0,2) stability
diagram transition. Bottom: overlap of the $\Psi$ state with
$S(0,2)$ (a), with $\bket$ (b) and with $\cket$ (c). Parameters:
$B_{ext}=0.2 \, T$, $\gamma = 1.2 \, \mu eV$, $\Delta = 1000$,
$E_C=0.6 meV$, $V_x=1 \mu eV$.}
\end{figure}

Here we describe a force toward either equalizing or inducing
differences between the Overhauser fields in the two dots. The
direction of this force depends on the spin of the initial electron
state (i.e. the direction of the electron-nuclear ``flip-flop"
process) as well as on the sign of the product of the g-factor $g$
of the host material and the sign of the Overhauser magnetic field
of the nuclei of Ga and As as compared to the direction of the
nuclear spin (they are anti-parallel). Assuming GaAs, we describe
two pulse sequences which differ in the choice of the initial
electron state, which consequently have a force tending to cause the
Overhauser fields in the two dots to equilibrate or to diverge.

\emph{Electronic States of the Double Dot with N=2} - We calculate
the electronic states of the two electron ($N=2$) double dot within
the Hund-Mulliken formalism \cite{Burkard05} developed for the
hydrogen molecule. Typically, in this method, eigenstates of total
spin, singlets and triplets, are employed as basis states. However,
since we wish to study the inhomogeneous Overhauser effect due to
different effective magnetic fields in the two dots, we choose a
basis which diagonalizes, at the single particle level, the
z-component of this inhomogeneous field and in which the spatial
dependence of nuclear spin flips induced by electronic spin ``flops"
is transparent. The basis is: $\{\xi_n\} \equiv \{\rrket, \, \bket,
\, \cket, \, \tket \}$, where L and R indicate the orbital states of
the left and right dot, the arrows denote spin direction
\cite{standard}. Two remaining states of the Hund-Mulliken model,
$\makeket{L}{\da}{R}{\da}$ and $\makeket{L}{\ua}{L}{\da}$ are not
relevant to our analysis. Note that $\rrket$ is the standard $\s$
state and $\tket$ is the standard $\Tp$ state. The hyperfine
Hamiltonian for two electrons is properly written:
\begin{equation}
H_{hf} = \frac{v\hftot}{\hbar^2} \sum_m^M [\delta({\bf r}_1-{\bf
R}_m) {\bf S}_1 \cdot {\bf I}_m \otimes {\bf 1} + {\bf 1} \otimes
\delta({\bf r}_2-{\bf R}_m) {\bf S}_2 \cdot {\bf I}_m] \label{eq:hf}
\end{equation}
where ${\bf r}_i$ and ${\bf S}_i$ are operators in the subspace of
electron $i$ (first quantized representation) and $m$ is summed over
a total of $M$ nuclei (typically $M \sim 10^6$); and where $v$ is
the volume per nucleus. We assume, for simplicity, a single nuclear
species with spin $1/2$. Then, constraining the maximum Overhauser
field to be $5.3 \, T$ \cite{Taylor07} leads to a coupling constant
$\hftot =- 270 \, \mu eV$. We incorporate the matrix elements of
$H_{hf}$ from Eq. \ref{eq:hf} in our basis $\{\xi_n\}$ into the
Hund-Mulliken Hamiltonian which gives
\begin{widetext}
\begin{equation}
\begin{split}
& \hspace{0.9cm} \rrket \hspace{1.2cm} \bket \hspace{1.9cm} \cket
\hspace{2.5cm} \tket \\
 H &= \begin{pmatrix} E_C - \varepsilon \hspace{1.0 cm} & I_z^{LR} -
I_z^{RR}\LR + \gamma
& I_z^{LR} - I_z^{RR}\LR + \gamma  & I_+^{RR}\LR - I_+^{LR} \\
& I_z^{LL} - I_z^{RR}
 & V_x & I_+^{RR} \\
& & I_z^{RR} - I_z^{LL}
& I_+^{LL}  \\
& & &  I_z^{RR} + I_z^{LL} +E_Z \label{eq:H}
\end{pmatrix},
\end{split}
\end{equation}
\end{widetext}
which is correct to leading order in $\LR$,
where $\varepsilon$ is the potential ``detuning'' (the difference
between the electrostatic potential bottom of the left and right
dots) and where we have taken the orbital energies of $L$ and $R$ to
be zero for simplicity. We include only two Coulomb terms: the
charging energy $E_C \equiv V_{RRRR} - V_{RLRL}$ and the exchange
matrix element $V_x \equiv V_{LRRL}$ \cite{me}. Also in Eq.
\ref{eq:H}, $\gamma$ is the tunneling coefficient; $E_Z \equiv g
\mu_B B_{ext}$ is the Zeeman energy for a presumed external field
$B_{ext}$, which defines the z-axis of the problem, with $\mu_B$ the
Bohr magneton. The lower left hand side of the
matrix is the complex conjugate of the upper right hand side. Note
that the matrix elements of $H$ in this electronic basis remain
operators in the Hilbert space of the nuclear coordinates
\cite{identity}:
\begin{equation}
\vec{I}^{\alpha \beta} \equiv v \frac{\hftot}{2 \hbar}
\sum_{m=1}^{M} \psi_\alpha^*({\bf R}_m) \psi_\beta({\bf R}_m)
\vec{I}_{m}
\end{equation}
where $\alpha,\beta \in \{L,R\}$. Note that previous researchers
have typically ignored the transition term $I_{+}^{LR}$, which we
see from Eq. \ref{eq:H} can lead to a direct transition between
$\rrket$ and $\tket$ and causes a spin flip \emph{in the barrier}.
This term, and the other overlap terms (e.g. $\propto \langle L|R
\rangle$), could be significant in the case of large $B_{ext}$ and
small $\gamma$, i.e. where $\rrket$ and $\tket$ anti-cross deep in
the (0,2) regime. However we will henceforth ignore hyperfine terms
in Eq. \ref{eq:H} proportional to wavefunction overlap (e.g. $
\langle L|R \rangle $ and $I_+^{LR}$) and leave exploration of the
barrier nuclear spin flip to a later publication \cite{Stopa09}.

\emph{Nuclear spin flip location} - The crucial feature of Eq.
\ref{eq:H} is that the $\tket$ state is coupled to $\bket$ via a
term which flips a nuclear spin in the right dot ($I_+^{RR}$) and it
is coupled to $\cket$ by a term that flips a nuclear spin in the
left dot ($I_+^{LL}$) . In the \emph{absence} of flip-flop coupling
to the $\tket$ state, the upper left 3x3 matrix in Eq. \ref{eq:H}
(see also yellow highlighted region of Fig. 2) has a ground state,
which we denote:
\begin{equation}\label{eq:psi}
\Pket = a(\varepsilon) \rrket + b(\varepsilon) \bket +
c(\varepsilon) \cket.
\end{equation}
As shown in figure 1, at large (positive) $\varepsilon$, $\Pket
\rightarrow \rrket \equiv S(0,2)$ and at large negative
$\varepsilon$, $\Pket$ becomes an unequal superposition of $\bket$
and $\cket$. Even when $V_x > |\langle I_z^{RR}-I_z^{LL} \rangle |$,
the inhomogeneous Overhauser effect will produce a preference for
either the $\bket$ or the $\cket$ component of $\Psi$ (see figure
1), with the electron down spin preferentially located on the dot
with smaller $I_z$. In the first electron pulsing sequence which we
describe, the electron state is initialized at large $\varepsilon$
into $\Pket \approx S(0,2)$ and detuning is swept approximately
adiabatically through the $\Psi$ - $\tket$ anti-crossing. The
position of this anti-crossing, $\tilde{\varepsilon}$, is determined
by the energy of $\tket \equiv \Tp$, denoted $E(T+)$ (see figure 1),
which is determined by $B_{ext}$. Insofar as $b(\tilde{\varepsilon})
\ne c(\tilde{\varepsilon})$, a transition from $\Psi$ to $\tket$
will preferentially induce a nuclear spin flip (down) on the side
with the larger $I_z$. This tends to equilibrate the values of
$I_z^{RR}$ and $I_z^{LL}$. In the second pulse sequence the
electrons are initialized into $\tket$ and the state then
transitions to $\Psi$. The same feedback mechanism preferentially
now causes nuclear spins to flip \emph{up}, but still on the side
with the larger $I_z$, thus leading to a tendency for $|I_z^{LL} -
I_z^{RR}|$ to grow. Both of these sequences can be experimentally
implemented \cite{Petta05,Foletti09}. Our further analysis focuses
mainly on the first pulse sequence.

Note that the preceding argument depends on the \emph{sign} of $\hftot$
which in turn depends on the sign of $g$ and the sign of the
effective Overhauser field which, for Ga and As, are anti-parallel
to the nuclear spins \cite{Paget77}.

\emph{Nuclear States} -To further analyze the Hamiltonian, Eq.
\ref{eq:H}, it is helpful to introduce a simplified basis for the
nuclear states in which all of the nuclei are either in the left or
right dot and all within a given dot interact equally with the
electron. In other words, $|\psi_L(\br)|^2$ is taken as a constant
within a spherical ``box" of some volume, $\cal V$. In this model,
which we refer to as the ``box model,'' the squares of the total
angular momenta $I_{\alpha}^2$ are conserved, where
$\vec{I}_{\alpha} \equiv (v \hftot /{\cal V}) \sum_{m \in \alpha}
\vec{I}_m$, and where $\alpha \in \{L,R\}$. Thus, the electrons
essentially interact with two composite nuclear spins, one on the
left and one on the right. The nuclear state basis is
$\{I_L,I_R,I_{Lz},I_{Rz}\}$ (where $I_{\alpha} (I_{\alpha} +1)$ is
the eigenvalue of $(\vec{I}^{\alpha \alpha})^2$ and $I_{\alpha z}$
is the eigenvalue of $I_z^{\alpha \alpha}$). Finally, for given
$I_L, I_R$, it is convenient to transform to the basis of $\Delta
\equiv I_{Lz} - I_{Rz}$ and $s \equiv I_{Lz} + I_{Rz}$. In this
basis the z-components of the nuclear operators have non-zero matrix
elements on the diagonal blocks, but the raising and lowering
operators connect different $(\Delta,s)$ subspaces (see Fig. 2).

The strength of the narrowing force depends on the ratio $r \equiv
c/b$ at $\tilde{\varepsilon}$. This depends on $\Delta$ and on
$B_{ext}$. For example, smaller $B_{ext}$ results in smaller (or
more negative) $\tilde{\varepsilon}$, where, as shown in Fig. 1, the
ratio $c/b$ increases (for $\Delta > 0$). Exactly \emph{how} large
$c/b$ can get depends on $V_x$ which, in the example of Fig. 1, we
have set to $1$~$\mu$eV \cite{vx}.

In Fig. 3 we plot the value of $r(\tilde{\varepsilon})$ as a
function of $B_{ext}$ for various values of $\Delta$. The
key point is that $r(\tilde{\varepsilon})$ increases monotonically
with $\Delta$ (cf. yellow highlighted region of Fig. 2), however it
also decreases monotonically with $B_{ext}$ (and hence
$\tilde{\varepsilon}$). Interestingly,
because the $\rrket$ state is coupled equally to $\bket$ and
$\cket$, the value of $b/c$ is \emph{independent of $\gamma$}.

We note that the flip-flop process naturally also depends on the
rate at which $\varepsilon$ is swept since, in order to be adiabatic
and remain on the lower branch of the $\Psi$ - $\tket$ anti-crossing
the $\varepsilon$ variation must be sufficiently slow. More
generally, the character of the state evolution can be examined as a
Landau-Zener tunneling problem \cite{Gullans09} or else evaluated
numerically \cite{Stopa09}.

\begin{figure}
\begin{center}
\includegraphics[bb=0.0in 1.0in 13.0in 13.0in,width=3.375in]{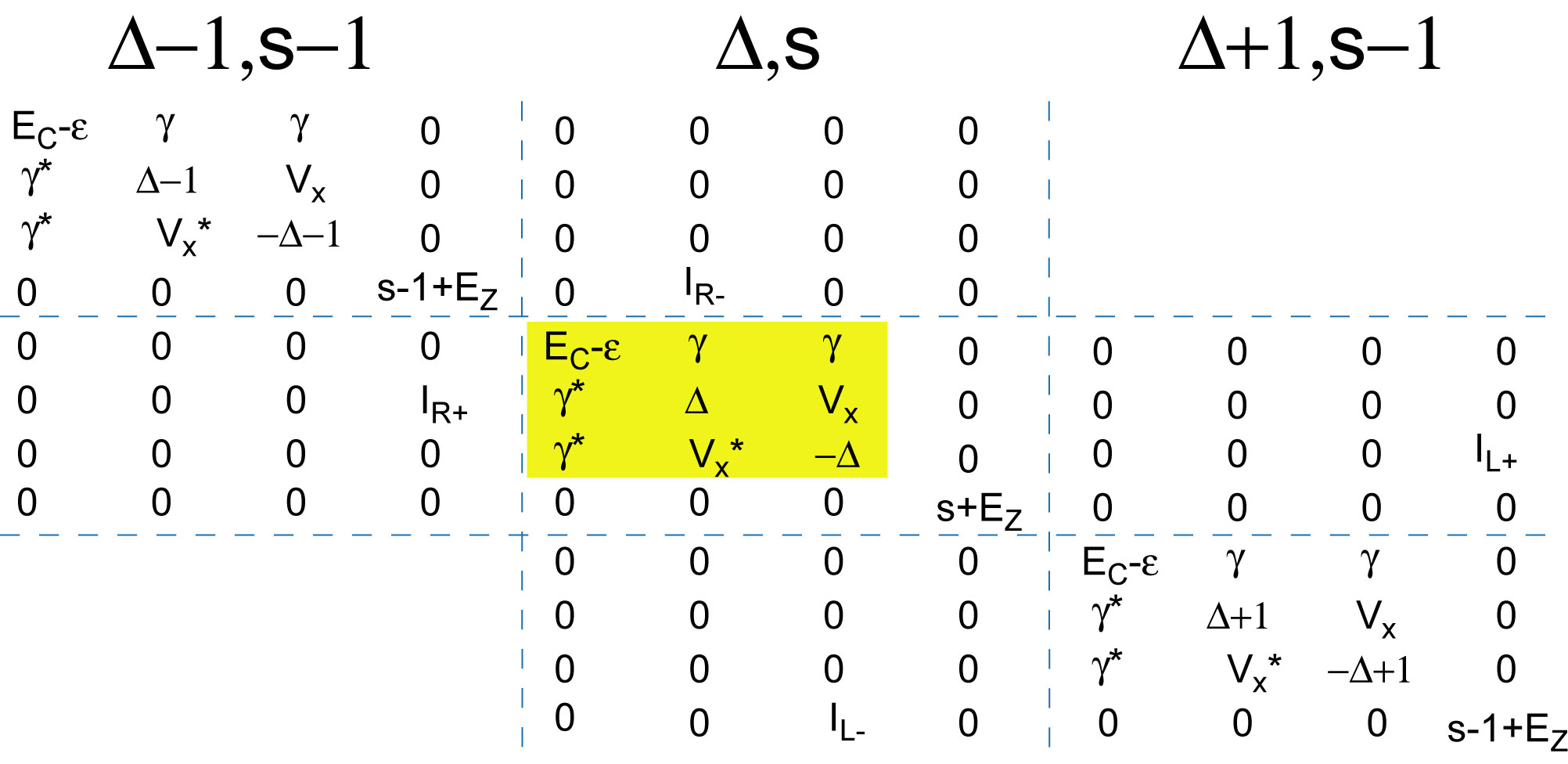}
\end{center}
\caption{Hamiltonian for three sectors of the nuclear difference
quantum number $(\Delta-1,s-1),(\Delta,s),(\Delta+1,s-1)$. In the
above, $\pm \Delta \pm 1$ and $s \pm 1$ are shorthand for $(v \hftot
/{\cal V}) (\pm \Delta \pm 1)$ and $(v \hftot /{\cal V}) (s \pm 1)$
respectively.}
\end{figure}

The evolution of the full nuclear state is complex and the
experimental manifestations of that evolution are ambiguous. Nevertheless, as a possible baseline for more detailed studies of the nuclear evolution, we describe a simple, incoherent model which results in narrowing of the distribution of $\Delta$.

\begin{figure}
\begin{center}
\includegraphics[bb=0.0in 1.0in 11.0in 11.0in,width=3.375in]{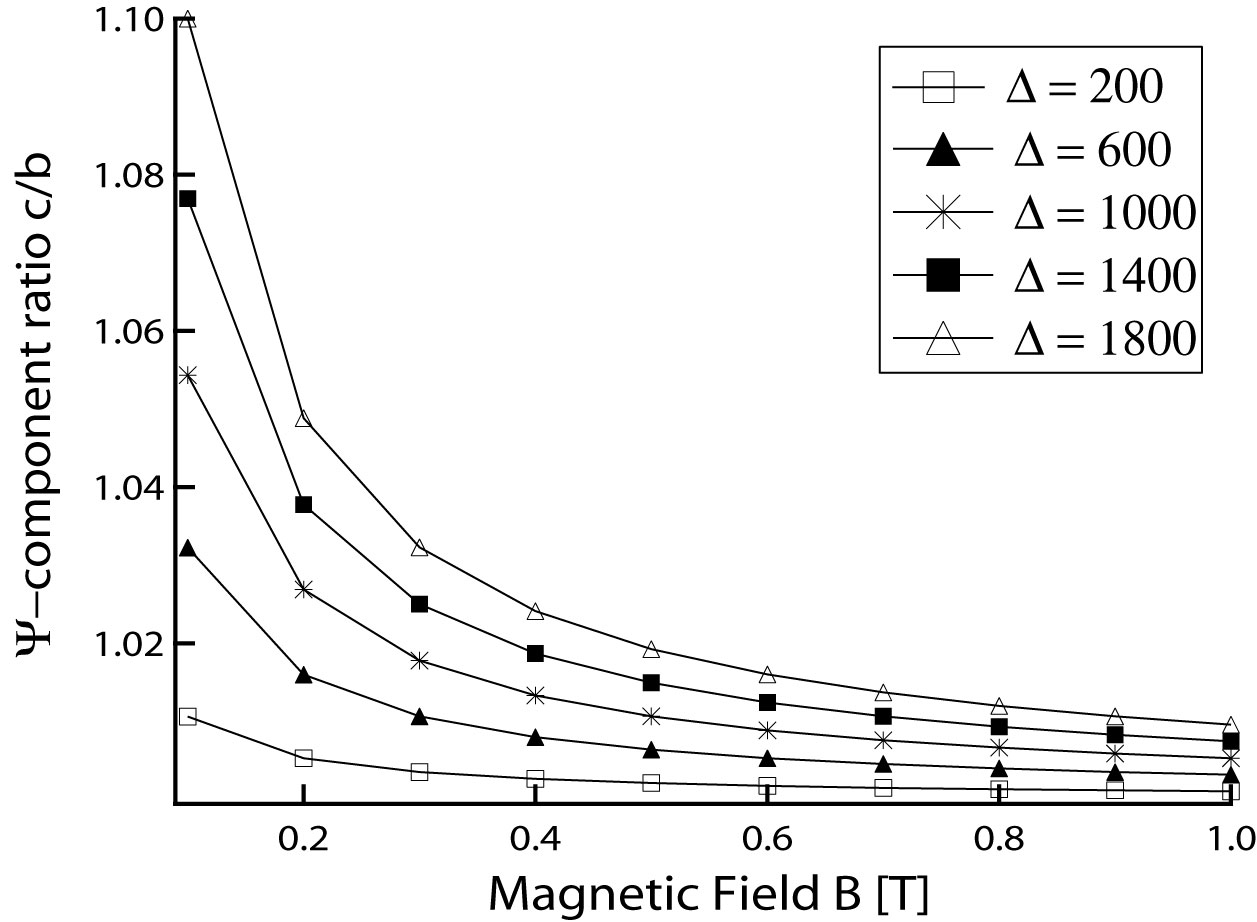}
\end{center}
\caption{The wave function ratio $r \equiv c/b$ evaluated at the
$\Psi - T+$ crossing point, $\tilde{\varepsilon}$, as a function of
$B_{ext}$ for various values of the nuclear spin z-component
difference $\Delta$. $r(\tilde{\varepsilon})$ is monotonically
increasing with $\Delta$ and decreasing with $B_{ext}$.}
\end{figure}

If we assume that the system is in the well-defined state $\Pket
\otimes \{I_L,I_R,I_{Lz},I_{Rz}\}$ and the detuning is moved quickly
to $\tilde{\varepsilon}$ and held there for time $\tau$, we can
compute, by Fermi's golden rule, the probability for a nuclear spin
to flip in the right dot as:
\begin{equation}
\begin{split}
\Gamma_R(I_{Lz}, I_{Rz} & \rightarrow I_{Lz}, I_{Rz} - 1) \equiv
\Gamma_R(s, \Delta
\rightarrow s-1, \Delta-1) \\
=& \frac{\tau^2}{\hbar^2} | \langle I_{Lz} I_{Rz}-1
|I_-^{RR}|I_{Lz} I_{Rz} \rangle |^2 \\
=& \frac{\hftot^2 \Omega_{R-}^2 \tau^2}{4 \hbar^2}  |b|^2
\end{split}
\end{equation}
where we have suppressed the $I_L, I_R$ dependence for brevity and
where the matrix elements of the ladder operators are given by the
well-known formulas: $\Omega_{\alpha \pm} \equiv \langle I_{\alpha},
I_{\alpha z} \pm 1 | I_{\pm} | I_{\alpha}, I_{\alpha z} \rangle =
\sqrt{I_{\alpha}(I_{\alpha}+1) - I_{\alpha z}(I_{\alpha z} \pm 1)}$.
Similarly, the flip probability in the left dot is proportional to
the $c$ component of $\Psi$
\begin{equation}
\Gamma_L (s, \Delta\rightarrow s-1, \Delta+1)= \frac{\hftot^2
\Omega_{L-}^2 \tau^2}{4 \hbar^2} |c|^2.
\end{equation}
If we denote the probability distribution for the nuclear state (at
fixed $I_L,I_R$) as $W(s, \Delta)$, then the condition for W to be
stable in its dependence on $\Delta$ can be written (cf. Fig. 4a):
\begin{equation}
\begin{split}
W(s+1, \Delta+1) & \Gamma_L(s+1,  \Delta+1)=W(s,\Delta) \Gamma_R
(s, \Delta)\\
W(s,\Delta + 1)&=W(s,\Delta) \frac{\Omega_{R-}^2(s, \Delta)
}{\Omega_{L-}^2(s, \Delta+1)}\frac{|b(\Delta)|^2}{|c(\Delta+1)|^2}
\label{eq:W}
\end{split}
\end{equation}
where we have assumed that $W(s) \approx W(s+1)$ and we have used
the fact that $b$ and $c$ depend very weakly on $s$ (only through
the $s$-dependence of $\tilde{\varepsilon}$).

Recursion relation Eq. \ref{eq:W} can be solved iteratively and the
influence of the narrowing force evaluated. In Fig. 4 we have
plotted $W(\Delta)$ computed with the ratio
$\Omega_{R-}/\Omega_{L-}$ set to unity to show only the narrowing
from the inhomogeneous Overhauser effect described here with the
same electronic parameters as in Fig. 1, and with $I_L = I_R =
1000$; including the $\Omega$'s induces more narrowing. For
comparison we show the $T \rightarrow \infty$ thermal distribution
of $\Delta$, averaged over $s$, also for $I_L = I_R = 1000$. Inset
(a) shows the ratio of the root-mean-square (rms) $\Delta$ in the
thermal distribution, $\sigma_T$, to the rms $\Delta$ with the
narrowing force at varying $B_{ext}$, $\sigma(B_{ext})$. A
substantial narrowing of $W(\Delta)$ results from the inhomogeneous
Overhauser effect.

\begin{figure}
\begin{center}
\includegraphics[bb=0.0in 1.0in 10.0in 10.0in,width=3.375in]{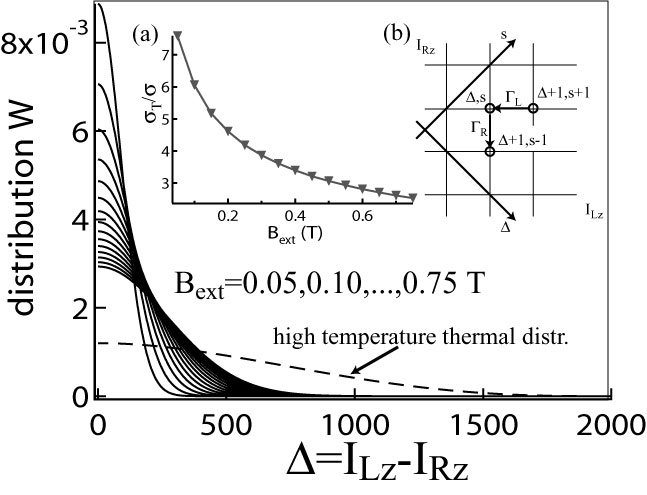}
\end{center}
\caption{(main) Reduced distribution $W(\Delta,s)$, calculated from Eq.
\ref{eq:W} (solid lines), for $V_x=1$~$\mu$eV as a function of $\Delta$ for various
$B_{ext}=0.05,0.10,...,0.75 \, $T (lower fields have narrower $W$);
and thermal $W(\Delta)$ (dashed), averaged over $s$, all with $I_L = I_R
=10^3$. Inset (a) narrowing factor $\sigma_T/\sigma(B_{ext})$ versus
$B_{ext}$. Inset (b) Illustration of $I_{Lz}-I_{Rz}$ plane. $\Delta$
and $s$ are the diagonal coordinates, with $\Delta \equiv I_{Lz} -
I_{Rz}$.}
\end{figure}

\emph{Discussion} - Experimentally, the $\Pket$ to $\tket$ pulse
sequence polarizes only about 1\% of the nuclei, even when running
sufficient cycles to flip all of the nuclei \cite{Petta05}. This
saturation of the nuclear polarization is still an open problem. A
recent article by Yao \cite{Yao09} discusses a model similar to that
described herein. In that paper, no mechanism for stopping the
flip-flop process is proposed when the pumping continues (as it does
in experiments) beyond $\sim 10^5$ cycles. In our model,
polarization will saturate when both $I_{Lz} = -I_L$ and $I_{Rz} =
-I_R$, implying that $s = -I_L - I_R$. However, the resulting
distribution of $\Delta$ will then mirror the difference in the
initial distributions of $I_L$ and $I_R$, and hence will show no
narrowing of $W(\Delta)$. Thus our box model can qualitatively
explain the narrowing effect or the saturation, but not both.

We believe that a full understanding of these phenomena depends on
the variable coupling of the electron wave function to different
groups of nuclei, so that conservation of the magnitudes of two
spins, $\vec{I}_L$ and $\vec{I}_R$, is not required. Such a model
with multiple interacting composite nuclear spins, incorporating the
narrowing effect described here as well as the Landau-Zener
tunneling behavior near $\tilde{\varepsilon}$, in some parameter
regimes shows the potential to send $\abs{\Delta}\rightarrow0$ while
reducing the spin flip probability, slowing the growth of total
polarization; for other parameters, $\abs\Delta$ grows large despite
the narrowing force described here \cite{Gullans09}.

\emph{Acknowledgments} - We thank B. I. Halperin, M. Gullans, J.
Taylor, M. Lukin, S. Foletti, H. Bluhm, Y. Tokura and M. Rudner for
valuable conversations. We thank the National Nanotechnology
Infrastructure Network Computation Project for computational
support. We gratefully acknowledge support from the Fannie and John
Hertz Foundation, NSF grants PIF-0653336 and DMR-05-41988 and the
ARO.


\begin{references}
\bibitem{Marcus} A. C. Johnson \emph{et al.}, Nature, \textbf{435},
925 (2005).

\bibitem{Ono04} K. Ono and S. Tarucha, Phys. Rev. Lett. \textbf{92},
256803 (2004).

\bibitem{Loss98} D. Loss and D. DiVincenzo, Phys. Rev. A
\textbf{57}, 120 (1998).

\bibitem{DasSarma06} W. M. Witzel and S. Das Sarma Phys. Rev. B
\textbf{74}, 035322 (2006).

\bibitem{Vandersypen04} See, for example, L. M. K. Vandersypen and I. L. Chuang, Rev.
Mod. Phys. \textbf{76}, 1037 (2004), and references therein.

\bibitem{Hanson07} R. Hanson \emph{et al.}, Rev. Mod. Phys.
\textbf{79}, 1217 (2007).

\bibitem{Ono02} K. Ono \emph{et al.}, Science \textbf{297}, 1313
(2002).

\bibitem{Reilly08} D. Reilly \emph{et al.}, Science \textbf{321},
781 (2008).

\bibitem{Ramon_Hu07} G. Ramon and X. Hu, Phys. Rev. B \textbf{75},
161301 (2007).

\bibitem{Ribeiro08} H. Ribeiro and G. Burkard, arXiv:0811.3560v1


\bibitem{Foletti09} S. Foletti, H. Blum, D. Mahalu, V. Umansky and
A. Yacoby, in preparation.

\bibitem{Burkard05} G. Burkard, D. Loss and D. DiVincenzo, Phys.
Rev. B, \textbf{59}, 2070 (1999).

\bibitem{standard} More traditionally, the linear
combinations $S(1,1) \equiv [\bket - \cket]/\sqrt{2}$ and $T_0 (1,1)
\equiv [\bket + \cket]/\sqrt{2}$, are employed.

\bibitem{Taylor07} J. M. Taylor, \emph{et al.}, Phys. Rev. B
\textbf{76}, 035315 (2007).

\bibitem{me} Coulomb matrix elements are defined in the usual way in
our two state basis, $\alpha,\beta,\gamma,\delta \in \{L,R\}$:
$V_{\alpha \beta \gamma \delta} \equiv \int \int \, d \br_1 d \br_2
\psi_\alpha^* (\br_1) \psi_\beta^* (\br_2) V(\br_1, \br_2)
\psi_\gamma (\br_1) \psi_\delta(\br_2)$.

\bibitem{identity} We have also used the identity: ${\bf S} \cdot {\bf I}_m =
S_z I_{mz}+ [S_- I_{m+} + S_+ I_{m-}]/2$.


\bibitem{Stopa09} M. Stopa, unpublished.

\bibitem{Petta05} J. R. Petta \emph{et al.}, Science \textbf{309},
2180 (2005).

\bibitem{Paget77} D. Paget \emph{et al.}, Phys. Rev. B \textbf{15}, 5780 (1977).

\bibitem{vx} Self consistent electronic structure calculations show
(Stopa, unpublished) that for lateral double quantum dots, $V_x$ can
range from $250 \, \mu V$ to less than $1\, \mu V$. Here, we have
chosen the lower value.

\bibitem{Gullans09} M. Gullans \emph{et al.}, in preparation.



\bibitem{Yao09} W. Yao arXiv:0905.2460v1


\end{references}
\end{document}